\documentclass[a4paper,12pt]{article}
\usepackage[utf8]{inputenc}
\usepackage{amssymb}
\usepackage[english]{babel}
\usepackage{amsthm}
\usepackage{mathrsfs}
\usepackage{hyperref}
\usepackage{dirtytalk}
\usepackage{physics}
\usepackage{xcolor}
\usepackage{stmaryrd}
\usepackage{yfonts}
\usepackage{graphicx}
\usepackage{mathtools}
\usepackage{subcaption}
\usepackage{chngcntr}
\counterwithin{figure}{section}
\counterwithin{table}{section}
\usepackage{caption}
\usepackage{floatrow}
\usepackage{courier}
\usepackage{cases}
\usepackage{array}
\usepackage{booktabs}
\usepackage{amsmath}
\usepackage{amsmath, amssymb, graphics, setspace}
\usepackage{titling}
\usepackage[tmargin=3cm, bmargin=3cm,inner=3cm,outer=3cm]{geometry}
\usepackage{authblk}
\usepackage{url}

\title{\bf Quantum Statistical Mechanics of Dissolving Vortices}
\author[$\textasteriskcentered$]{\Large N.S. Manton}
\author[$\dagger$]{\Large Shiyi Wang}

\affil[$\textasteriskcentered$]{Department of Applied Mathematics and Theoretical Physics, \

Wilberforce Road, Cambridge CB3 0WA, U.K.}
\affil[$\dagger$]{St John's College, Cambridge CB2 1TP, U.K.}
\date{}
\begin{document}
\maketitle
\begin{abstract}
The quantum partition function for dissolving Abelian Higgs vortices
is calculated explicitly, using spectral data for the Beltrami Laplacian
on the $N$-vortex moduli space $\mathbb{CP}^N$ with a scaled Fubini--Study
metric. From the partition function, the pressure of the vortex gas
is derived. There are three asymptotic regimes -- \textit{High},
\textit{Intermediate} and \textit{Low Temperature}. The phase crossover
from \textit{Intermediate} to \textit{Low Temperature} is modelled
by a Bessel function. In the \textit{Low Temperature} regime the
free energy is not extensive but is proportional to $N^2$.
\end{abstract}

\section{Introduction}

In this paper we will consider the quantum statistical mechanics of
vortices, where by vortices we mean critically coupled Abelian Higgs
vortices on a compact, oriented surface $\Sigma$ having a fixed
Riemannian metric with area $A$. On such a surface,
there are static $N$-vortex solutions satisfying first-order
Bogomolny equations, provided the Bradlow inequality $4\pi N < A$ is
satisfied~\cite{Bra}. The vortex number $N$ is a topological
invariant, and a vortex can be informally seen as a `hard disk' of
area $4\pi$. Here we assume $N$ is large, and correspondingly $A$
sufficiently large, so that we can discuss the statistical mechanics
of vortices.

The moduli space $\mathcal{M}_N$ of $N$-vortex solutions has real
dimension $2N$ because the vortex centres, where the Higgs field
vanishes, can be at arbitrary locations on $\Sigma$.
Vortices are indistinguishable, so $\mathcal{M}_N$ is
diffeomorphic to $\Sigma^N/S_N$, the $N$th symmetrised power of
$\Sigma$~\cite{book}. We will assume here that
$\Sigma$ is diffeomorphic to $S^2 = \mathbb{CP}^1$.
$\mathcal{M}_N$ is then diffeomorphic to $\mathbb{CP}^N$.

The moduli space $\mathcal{M}_N$ acquires from the kinetic part of the
Lagrangian of the Abelian Higgs model a Riemannian (in fact K\"ahler)
metric that depends on the metric of $\Sigma$. In general, the moduli space
metric is not known explicitly, but some of its global properties,
in particular its volume \cite{MN} and total scalar curvature
\cite{Bap}, are known. Classical $N$-vortex dynamics can be approximated
by geodesic motion on $\mathcal{M}_N$ with this metric. Analogously,
we shall assume here that quantized vortex dynamics is modelled by
a Hamiltonian proportional to the Beltrami Laplacian on $\mathcal{M}_N$. 

The classical statistical mechanics of $N$ vortices on
$S^2$ with a round metric of area $A$ is well-understood using only the
volume of $\mathcal{M}_N$ \cite{Man}. The vortices behave as
an interacting gas of finite-sized particles, where the pressure is
of the Clausius form
\begin{equation}
P = \frac{NT}{A-4\pi N} \,.
\label{Clausius}
\end{equation}
It was noted in~\cite{book} that this classical formula for the
pressure remains valid for vortices on surfaces $\Sigma$ with arbitrary
underlying metric and having any genus. The quantum statistical mechanics of
vortices is less fully understood, and almost certainly has some
dependence on the metric on $\Sigma$. In~\cite{Man22}, the first
quantum correction to the classical partition function at high
temperatures $T\gg\hbar^2$, and hence to the free energy and pressure,
was found. The correction depends on the total scalar curvature of
the moduli space $\mathcal{M}_N$. Because the temperature is high,
quantum effects are not dominant.

It is known that the metric on the moduli space $\mathcal{M}_N$
simplifies as one approaches the Bradlow limit, $A = 4\pi N$. The
regime where $A \gtrsim 4\pi N$ is called the regime of dissolving
vortices. Here the magnetic field approaches a uniform, constant value
and the Higgs field is small. (At the Bradlow limit itself, the magnetic
field is uniform and the Higgs field vanishes everywhere, so the
vortices have dissolved.) For $N$ dissolving vortices on a round
sphere, it was shown in \cite{BM} by approximate calculation that the moduli
space metric approaches the standard Fubini--Study metric on $\mathbb{CP}^N$,
scaled by $A - 4\pi N$. Recently, Garc\'ia Lara and Speight
have made this argument more rigorous \cite{GS}, and they have also shown
that the Fubini--Study metric arises for any metric on $\Sigma$,
provided $\Sigma$ is diffeomorphic to $S^2$. The moduli space
forgets the underlying metric in this limit. 

In this paper, we investigate the quantum statistical mechanics of the
vortex gas in this dissolving limit, 
exploiting the spectrum of the Beltrami Laplacian for the
Fubini--Study metric. We assume that $A$ is a fixed multiple of $N$,
slightly larger than $4\pi$, but there is no constraint on the
temperature $T$, so we can extend the earlier high temperature results
to lower temperatures. We calculate the quantum partition function
$Z$, and derive the free energy $F$
and pressure $P$ of the vortex gas. For general $T$, this requires
solution of a single transcendental equation, which is straightforward
numerically, and we also derive explicit formulae for $F$ and $P$ in
three asymptotic regimes: \textit{High}, \textit{Intermediate}
and \textit{Low Temperature}. The results are collected in eqs.(\ref{Freeall})
and (\ref{Pressureall}) below. Remarkably, a phase crossover
occurs between \textit{Intermediate} and \textit{Low Temperature}, and in the
\textit{Low Temperature} regime the free energy $F$ is no longer extensive,
but is proportional to $N^2$.

\section{Quantum Partition Function}

The classical Lagrangian for dynamics on the
$N$-vortex moduli space $\mathcal{M}_N$, using real coordinates $q^i$, is
\begin{equation}
L=\frac{1}{2}g_{ij}(\mathbf{q})\dot{q^i}\dot{q^j}
\end{equation}
where, for dissolving vortices, $g_{ij}=(A-4\pi N)G_{ij}$ with $G_{ij}$
the standard Fubini--Study metric on $\mathbb{CP}^N$~\cite{BM,GS}.
Passing to the Hamiltonian, we have
\begin{equation}
H=\frac{1}{2}g^{ij}(\mathbf{q})p_ip_j
\end{equation}
with $g^{ij}$ the inverse metric and $p_i$ the conjugate momentum to $q^i$.

The quantum Hamiltonian is taken to be
\begin{equation}
H=\frac{1}{2}\hbar^2\Delta
\end{equation}
where $\Delta=-\nabla^2$ is the Beltrami Laplacian
for the scaled Fubini--Study metric. Including the scale factor
$A-4\pi N$, the eigenvalues $\lambda_k$ of $\Delta$ and their
degeneracies $g_k$ are \cite{BGM}
\small
\begin{align}
\lambda_k&=\frac{4k(N+k)}{A-4\pi N} \,, \nonumber\\
g_k=\binom{N+k}{k}^2-\binom{N+k-1}{k-1}^2
  & =\left(1-\frac{k^2}{(N+k)^2}\right)\binom{N+k}{k}^2
  = \frac{N(N + 2k)}{(N+k)^2}\binom{N+k}{k}^2
\label{evalues}
\end{align}
\normalsize
where $k \in \mathbb{Z}^+$ and $\binom{N+k}{k}$ is the
combinatorial symbol. The quantum partition function $Z$ at
temperature $T$ is therefore
\begin{align}
Z(T) &= \sum_k g_k \exp\left(-\frac{\hbar^2}{2T} \lambda_k \right)
\nonumber \\
&=\sum_{k} \frac{N(N+2k)}{(N+k)^2} \binom{N+k}{k}^2
\exp\left(-\frac{\hbar^2}{2T}\frac{4k(N+k)}{A-4\pi N}\right)\nonumber \\
&\equiv \sum_{k} \frac{N(N+2k)}{(N+k)^2} \left(\frac{(N+k)!}{N! \, k!}\right)^2
\exp\left( -zk\left(1 + \frac{k}{N}\right)\right) = Z(z) \,.
\label{Partitionsum}
\end{align}
The key parameter here is the \textit{reciprocal temperature} $z$, defined as
\begin{equation}
z=\frac{\hbar^2}{2\pi T}\frac{4\pi N}{A-4\pi N} \,.
\label{reciptemp}
\end{equation}
We will use various techniques to calculate
$Z(z)$; the results are simpler when expressed in terms of $z$, but
are easily re-expressed in terms of $T$.

\section{High and Intermediate Temperatures}

It is well-known that sums like (\ref{Partitionsum}) may be approximated
for most $z$ by a Gaussian integral. Before doing this,
we simplify (\ref{Partitionsum}) using
Stirling's approximation $N! \simeq \sqrt{2\pi N}(\frac{N}{e})^N$ to obtain
\begin{equation}
Z(z) = \sum_k\frac{N + 2k}{N + k} \frac{1}{2\pi k}
\frac{(N+k)^{2(N+k)}}{N^{2N}k^{2k}}
\exp\left\{-zk\left(1+\frac{k}{N}\right)\right\} \,,
\label{ZStirl}
\end{equation}
and reorganise this in exponential form as
\begin{align}
  Z(z) &=\sum_k \exp \biggl(N\left\{
  2\left(1 + \frac{k}{N}\right)\log \left(1 + \frac{k}{N}\right)
  - 2 \frac{k}{N} \log \frac{k}{N}
  -z\frac{k}{N}\left(1 + \frac{k}{N}\right)\right\} \nonumber \\
  & \qquad\qquad\qquad + \log(N + 2k) - \log(N + k) - \log(2\pi k) \biggr) \,.
\label{Zexp}
\end{align}
The dominant terms are with $k=O(N)$ for $z\ll 2\log N$, as we will see,
so $\log Z(z)=O(N)$ and the $O(\log N)$ terms in the exponent
can be dropped. Hence,
\begin{equation}
  Z(z) \simeq \sum_k \exp \left(N \left\{
 2\left(1 + \frac{k}{N}\right)\log \left(1 + \frac{k}{N}\right)
  - 2 \frac{k}{N} \log \frac{k}{N}
  -z\frac{k}{N}\left(1 + \frac{k}{N}\right) \right\}\right) \,.      
\label{Zexpsimple}
\end{equation}
We now replace $\frac{k}{N}$ by $x$ and note that the spacing of the
$x$-values is $\frac{1}{N}$, so we can rewrite the sum as the integral
\begin{equation}
Z(z) \simeq N \int_0^\infty \exp (NG(x)) \, dx \,,
\label{ZG}
\end{equation}
where
\begin{equation}
G(x)=2(1+x)\log(1+x)-2x\log x-zx(1+x)
\label{Gdef}
\end{equation}
and the additional dependence of $G(x)$ on the reciprocal temperature $z$ is
implied.

$G(x)$ has a single maximum, with positive value; let us denote
its location by $x = x_0>0$. As $N$ is large, it is valid to
approximate (\ref{ZG}) as the Gaussian integral  
\begin{align}
Z(z) &\simeq N\int_{-\infty}^{\infty}\exp
\left(N\left\{ G(x_0) - \frac{G''(x_0)}{2}(x-x_0)^2 \right\}\right)\, dx
\nonumber \\       
&=\sqrt{\frac{2\pi N}{G''(x_0)}}\ \exp(NG(x_0))\simeq \exp(NG(x_0)) \,.
\label{ZGauss}
\end{align}
In the final expression we have again dropped a term
giving an $O(\log N)$ error in $\log Z(z)$. The free energy
$F$ of the vortex gas is therefore
\begin{equation}
F=-T\log Z(z)=-NTG(x_0) \,,
\label{Free}
\end{equation}
where the location $x_0$ of the Gaussian peak is related to the
reciprocal temperature $z$ via $G'(x_0)=0$. Explicitly, by differentiating
(\ref{Gdef}), we derive the relation
\begin{equation}
\log\left(1+\frac{1}{x_0}\right)=z\left(x_0+\frac{1}{2}\right) \,.
\label{zx_0rel}
\end{equation}

It is straightforward to numerically solve (\ref{zx_0rel}) to find
$x_0$ and use (\ref{Gdef}) to determine $G(x_0)$ and $F$ in terms of $z$, but
there are no exact formulae. Nevertheless, asymptotic formulae
for small and large $z$ can be obtained from the asymptotic solutions
of (\ref{zx_0rel}),
\begin{numcases}{x_0=}
  \frac{1}{\sqrt{z}}\left(1 + \frac{z}{24}
    + \frac{11z^2}{5760} +O(z^3)\right) -\frac{1}{2}
& \text{for small $z$} \label{x_0smallz}\\
e^{-\frac{1}{2}z}+(1-z)e^{-z}+O(e^{-\frac{3}{2}z}) & \text{for large
  $z$ \,,} \label{x_0largez}
\end{numcases}
and from these we obtain
\begin{numcases}
{F=}
-NT\left[1-\log
  z+\frac{z}{6}-\frac{z^2}{360}+O(z^3)\right]\qquad &
\text{for $0<z<5$} \label{Fsmallz}\\
-NT[2e^{-\frac{1}{2}z}+(1-z)e^{-z}+O(e^{-\frac{3}{2}z})] & \text{for
  $5<z\ll 2\log N$ \,.} \label{Flargez}
\end{numcases}
The ranges of validity will be discussed below.
As $z$ is the reciprocal temperature, eqs.(\ref{Fsmallz}) and
(\ref{Flargez}) correspond to the \textit{High Temperature} and
\textit{Intermediate Temperature}
regimes respectively. They extend the high temperature
result~\cite{Man22}
\begin{equation}
F=-NT\left[1-\log z+\frac{z}{6}+O(z^2)\right] \,,
\end{equation}
which holds for arbitrary underlying
surfaces\footnote{In~\cite{Man22} the term $\frac{z}{6}$
appeared incorrectly as $\frac{z}{12}$; this resulted from
misunderstanding factors of 2 in formulae for the total
scalar curvature.}. The formulae here are more precise and cover a
larger range of temperatures, but hold only for underlying
surfaces $\Sigma$ diffeomorphic to $S^2$ and vortex densities
$\frac{N}{A}$ close to the Bradlow density $\frac{1}{4\pi}$.

The \textit{Intermediate Temperature} asymptotic formulae have a limited
upper range of validity in $z$ because (\ref{x_0largez}) tells us
that the peak of the integrand in (\ref{ZG}) is at
$x_0\simeq e^{-\frac{z}{2}}$ for large $z$. So the dominant terms
in (\ref{Zexpsimple}) are at $k\simeq Ne^{-\frac{z}{2}}$,
implying that $k = O(N)$ precisely when $z\ll 2\log N$. Therefore,
(\ref{Flargez}) appears to be valid only for
$z\ll 2\log N$. In Section 4, we will see that its validity
essentially extends to $z<2\log N$, but the \textit{Low Temperature} regime
$z>2\log N$ needs to be considered separately.

\begin{figure}[H]
    \centering
    \includegraphics[width=15cm]{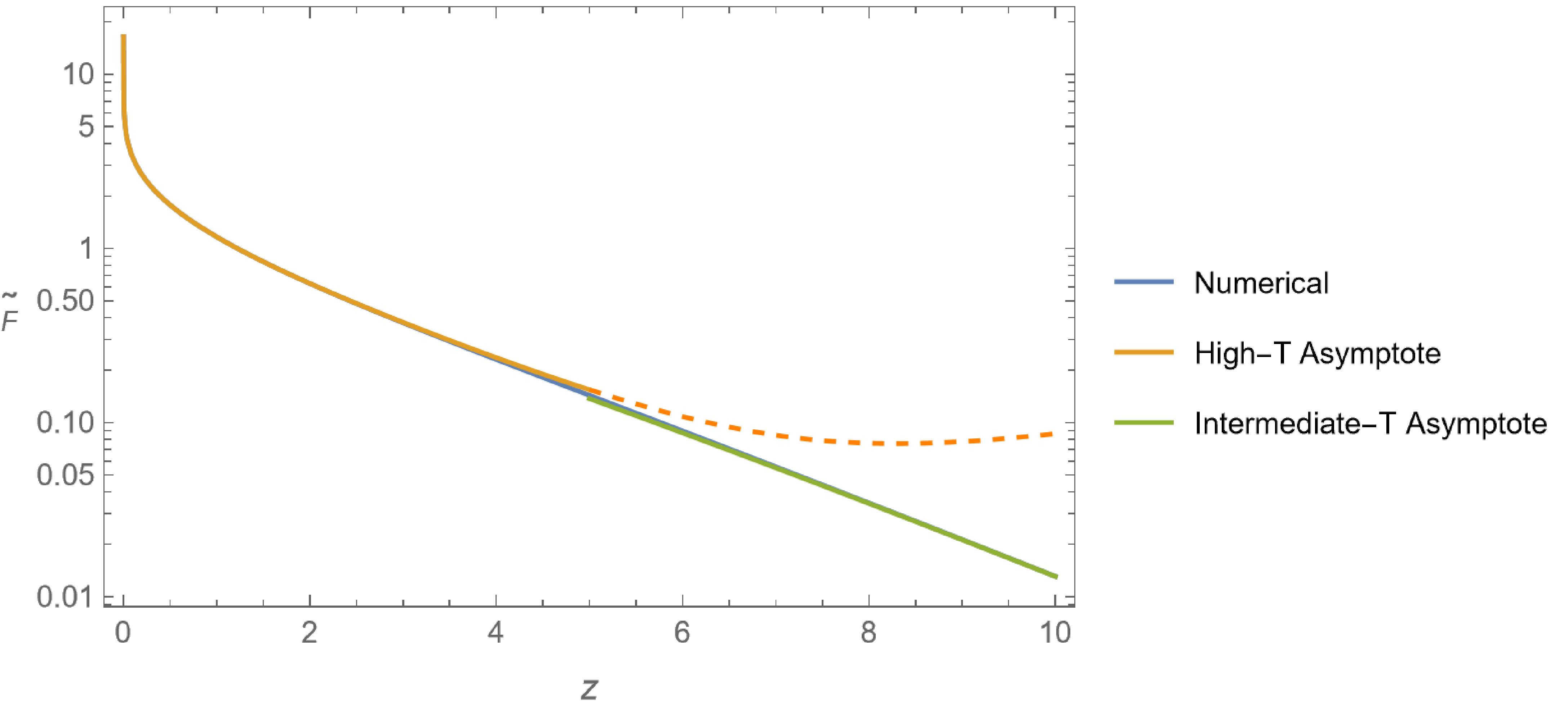}
    \caption{Reduced free energy $\widetilde{F}(z)$ for $z\ll 2\log N$.}
    \label{SmallF}
\end{figure}

Fig. \ref{SmallF} shows the \textit{reduced free energy}
$\widetilde{F}=F/(-NT)$, which is independent of $N$, as a
function of $z$. The asymptotic formulae work very well
-- the curve obtained using the numerical solution of
(\ref{zx_0rel}) is hardly visible. The \textit{High} and
\textit{Intermediate Temperature} asymptotics have a crossover
around $z=5$, which explains the ranges of 
(\ref{Fsmallz}) and (\ref{Flargez}).

We are mainly interested in the pressure of the gas,
obtained using the general expressions
\begin{align}
  P=-\frac{\partial F}{\partial A}
     = -\frac{dF}{dz}\frac{\partial z}{\partial A}
     =NT\frac{dG}{dz}\frac{\partial z}{\partial A}
     =NT\left(\frac{\partial G}{\partial z}
   +\frac{\partial G}{\partial x}
       \frac{dx}{dz}\right)\Bigg|_{x= x_0}
       \frac{\partial z}{\partial A} \,.
\end{align}
$x_0$ is the maximum of $G(x)$ for given $z$, so
$\frac{\partial G}{\partial x}\big|_{x= x_0} = 0$, and from formula
(\ref{Gdef}), $\frac{\partial G}{\partial z}\big|_{x= x_0} = -x_0(1+x_0)$.
Also, from (\ref{reciptemp}), $\frac{\partial z}{\partial A} =
-\frac{z}{A-4\pi N}$. Therefore   
\begin{equation}
P=\frac{NT}{A-4\pi N}[zx_0(1+x_0)] \,.
\label{Px_0z}
\end{equation}
Again, there is no exact formula for $P$ in terms of $z$. Nevertheless, the
following asymptotic formulae for small and large $z$ can be obtained
by using (\ref{x_0smallz}) and (\ref{x_0largez}),
\begin{numcases}
{P=}
\frac{NT}{A-4\pi N}
\left[1-\frac{z}{6}+\frac{z^2}{180}+O(z^3)\right]\qquad
& \text{for $0<z<5$} \label{PHigh} \\
\frac{NT}{A-4\pi N}
[ze^{-\frac{1}{2}z}+z(2-z)e^{-z}+O(e^{-\frac{3}{2}z})]
& \text{for $5<z\ll2\log N$} \,.
\label{PIntermediate}
\end{numcases}
These correspond to the \textit{High} and
\textit{Intermediate Temperature} regimes, and they extend the
high temperature result~\cite{Man22}
\begin{equation}
P=\frac{NT}{A-4\pi N}\left[1-\frac{z}{6}+O(z^2)\right] \,.
\end{equation}
We defer their plots to Section 4 where the formula in the
\textit{Low Temperature} regime $z>2\log N$ is also available.

\section{Phase Crossover and Low Temperature Regime}

Here, we derive the formulae that hold for $z>\log N$. They model a
phase crossover, from the \textit{Intermediate} to the
\textit{Low Temperature} regime.

\begin{figure}[H]
    \centering
    \includegraphics[width=15cm]{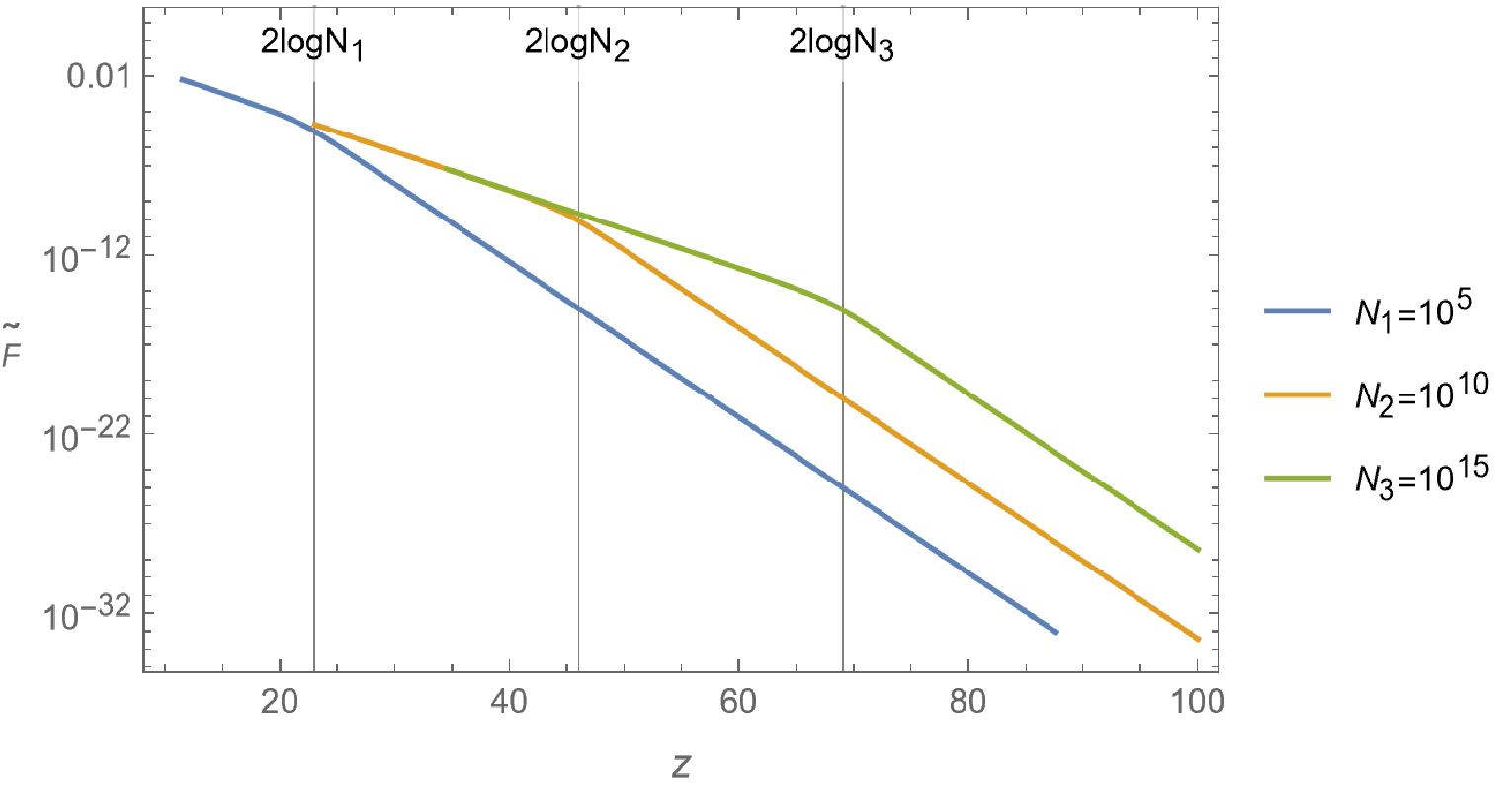}
    \caption{Reduced free energy $\widetilde{F}$ against reciprocal
      temperature $z$, showing the phase crossovers for $N=10^5$, $10^{10}$
      and $10^{15}$.}
    \label{FN}
\end{figure}

Recall the sum (\ref{Partitionsum}) defining the partition function
$Z(z)$. For large reciprocal temperature $z$, the sum is dominated
by terms with $k < \sqrt{N}$, so
\begin{equation}
Z(z) \simeq \sum_k \frac{N^{2k}}{(k!)^2} \exp(-zk) \,.
\label{Zlow}
\end{equation}
This is the series for the modified Bessel function
$I_0(2Ne^{-\frac{1}{2}z})$. One can show that precisely when
$z>\log N$, the terms in the series for $Z(z)$ and for
$I_0(2Ne^{-\frac{1}{2}z})$ are both negligible for $k > \sqrt{N}$.
So we can cut the tails and write
\begin{equation}
Z(z)\simeq I_0(2Ne^{-\frac{1}{2}z}) \,.
\label{ZBessel}
\end{equation}
The free energy of the vortex gas is then
\begin{equation}
F=-T\log Z(z)=-NT\left(\frac{1}{N}\log I_0(2Ne^{-\frac{1}{2}z})\right)
\qquad\text{for $z>\log N$} \,.
\label{FreeBessel}
\end{equation}
Fig. \ref{FN} shows the reduced free energy $\widetilde{F}=F/(-NT)$
on a logarithmic scale plotted against $z$ for $N=10^5$, $10^{10}$
and $10^{15}$. There is a rapid but smooth change in the slope for each $N$.
\noindent These phase crossovers take place approximately over the range
$2\log N - 3 < z < 2\log N + 3$. They require a delicate balance
between $z$ and $N$, so the finiteness of $N$ is crucial.
Note that for $N = 10^5$, $2\log N = 23.0 \gg 5$ so the
\textit{Intermediate} to \textit{Low Temperature} crossover is at much
lower temperature than the transition from the \textit{High} to
\textit{Intermediate Temperature} regime, and the crossover is
at an even lower temperature for larger $N$.

\begin{figure}[H]
\centering
\includegraphics[width=15cm]{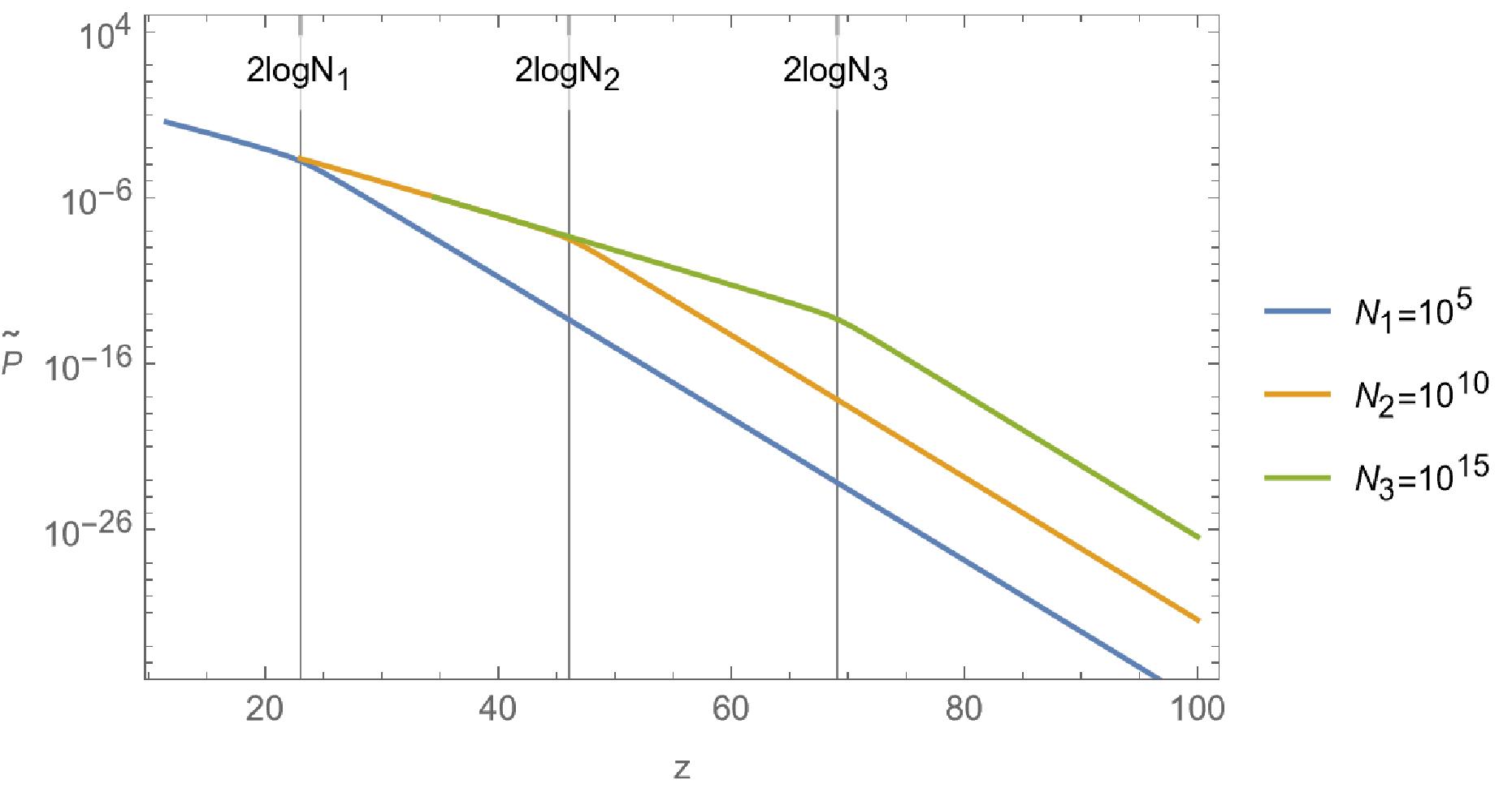}
\caption{Reduced pressure $\widetilde{P}$ against
reciprocal temperature $z$, showing the phase crossovers for $N=10^5$,
$10^{10}$ and $10^{15}$.}
\label{PN}
\end{figure}

There is a corresponding crossover in the pressure of the
gas. For $z > \log N$,
\begin{equation}
    P=-\frac{\partial F}{\partial A}=\frac{NT}{A-4\pi N}
    \left[\frac{I_1(2Ne^{-\frac{1}{2}z})}{I_0(2Ne^{-\frac{1}{2}z})} \,
    ze^{-\frac{1}{2}z}\right] \,,
\label{PBessel}
\end{equation}
where the Bessel function $I_1$ is the derivative of $I_0$. Since the
classical vortex gas has pressure
$\frac{NT}{A-4\pi N}$~\cite{Man}, it is natural to consider
the \textit{reduced pressure} $\widetilde{P}=P\,/\,(\frac{NT}{A-4\pi N})$,
the quantity in square brackets. This is shown in Fig. \ref{PN}.

We now clarify these results using \textit{Low Temperature} asymptotic
formulae. For $z>2\log N$, we can simplify the partition function
(\ref{ZBessel}) further. Introduce $\mu=Ne^{-\frac{1}{2}z}$, so
$Z(z) = I_0(2\mu)$ with $\mu<1$. The
series for $I_0(2\mu)$ has terms $1, \mu^2$, $\frac{1}{4}\mu^4$,
$\frac{1}{36}\mu^6$, $\dots$, which decrease super-geometrically.
Keeping just the leading terms gives
\begin{equation}
Z(z)=1+\mu^2+\frac{1}{4}\mu^4+\cdots \,.
\end{equation}
Using $\log(1+\lambda) = \lambda - \frac{1}{2} \lambda^2 + \cdots$,
the free energy and pressure in the \textit{Low Temperature} regime become
\begin{align}
F &=-T\log(1+\mu^2+\frac{1}{4}\mu^4+\cdots) \nonumber \\
  &=-T\left(\mu^2-\frac{1}{4}\mu^4+\cdots \right) \nonumber \\
  &=-NT\left[Ne^{-z} - \frac{1}{4}N^3e^{-2z} + \cdots \right] \,,
\label{FreeLow}
\end{align}
and
\begin{equation}
P=\frac{NT}{A-4\pi N}[Nze^{-z}-N^3ze^{-2z}+\cdots]
\qquad \text{for $z>2\log N$} \,.
\label{PLow}
\end{equation}
These \textit{Low Temperature} formulae are interesting.
For fixed $z$, the free energy $F$ is no longer extensive but is
proportional to $N^2$. This phenomenon violates classical
thermodynamics, which says $F$ should be proportional to $N$, but it
only occurs below some temperature $T$ inversely proportional to $\log N$.

In the region $z < 2\log N$ we can use the large-argument asymptotic
formulae
\begin{equation}
I_0(y) \sim \frac{e^y}{\sqrt{2\pi y}} \left(1 + \frac{1}{8y} + \cdots
\right) \,, \qquad
I_1(y) \sim \frac{e^y}{\sqrt{2\pi y}} \left(1 - \frac{3}{8y} + \cdots
\right)
\label{Bessellarge}
\end{equation}
to show that
\begin{equation}
P = \frac{NT}{A-4\pi N} \left[ ze^{-\frac{1}{2}z} - \frac{z}{4N} +
  \cdots \right] \,.
\label{BesselInterP}
\end{equation}
This matches the leading term in the pressure (\ref{PIntermediate}) in the
\textit{Intermediate Temperature} regime.

\begin{figure}[H]
  \centering
  \includegraphics[width=15cm]{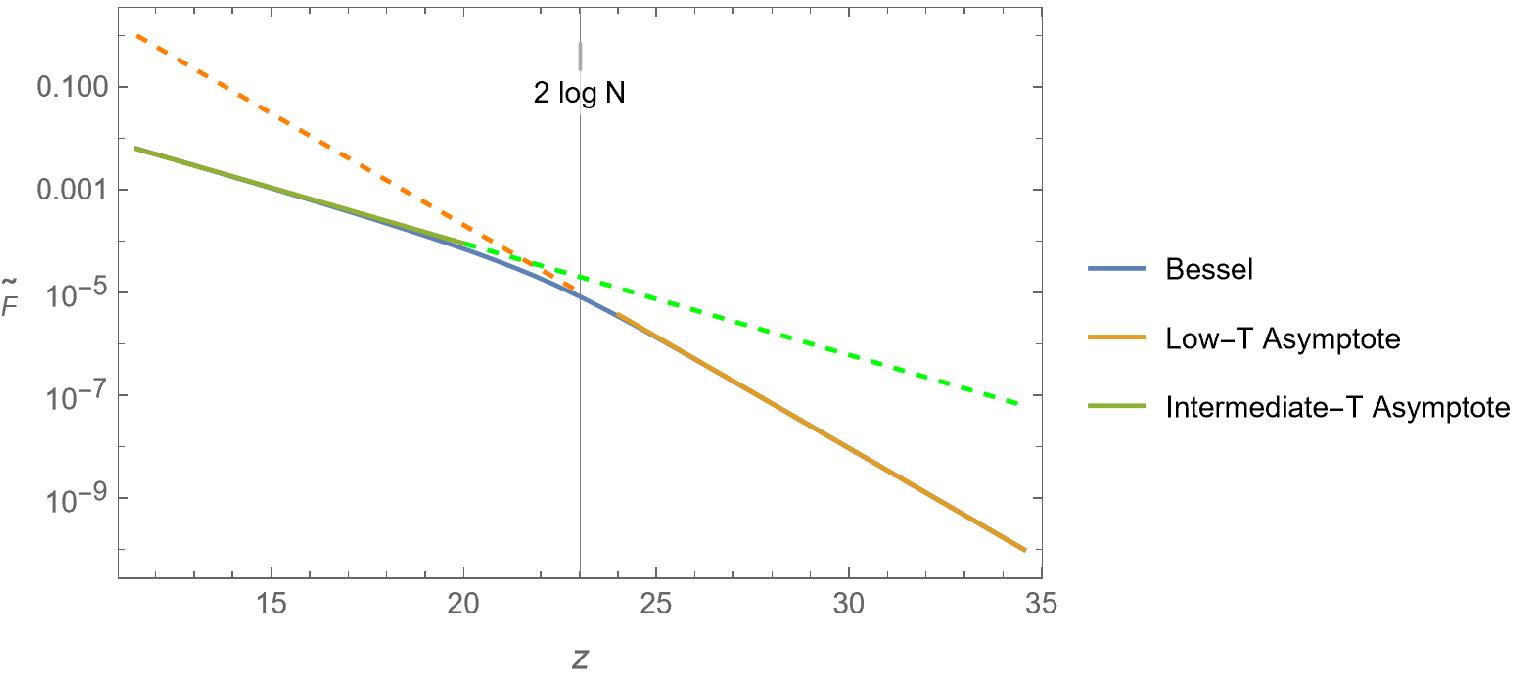}
  \caption{Reduced free energy $\widetilde{F}(z)$ (solid) and its
  asymptotes either side of the phase crossover for $N=10^5$.}
  \label{AsympF}
\end{figure}

The reduced free energy $\widetilde{F}$ for $N=10^5$ and $z>\log N$ is
shown in Fig. \ref{AsympF} together with the \textit{Intermediate Temperature} result,
previously known to be valid only for $z\ll 2\log N$. It is seen
that the \textit{Low Temperature} and \textit{Intermediate Temperature}
formulae are asymptotes of the more general, Bessel function result
derived from (\ref{FreeBessel}), which is shown solid but is hardly
visible away from $z \approx 2\log N$. In fact, these asymptotic
formulae cross over in the range $2\log N\pm 3$ for general $N$, so the
\textit{Intermediate Temperature} formula works almost up to $z=2\log N$.

\begin{figure}[H]
  \centering
  \includegraphics[width=15cm]{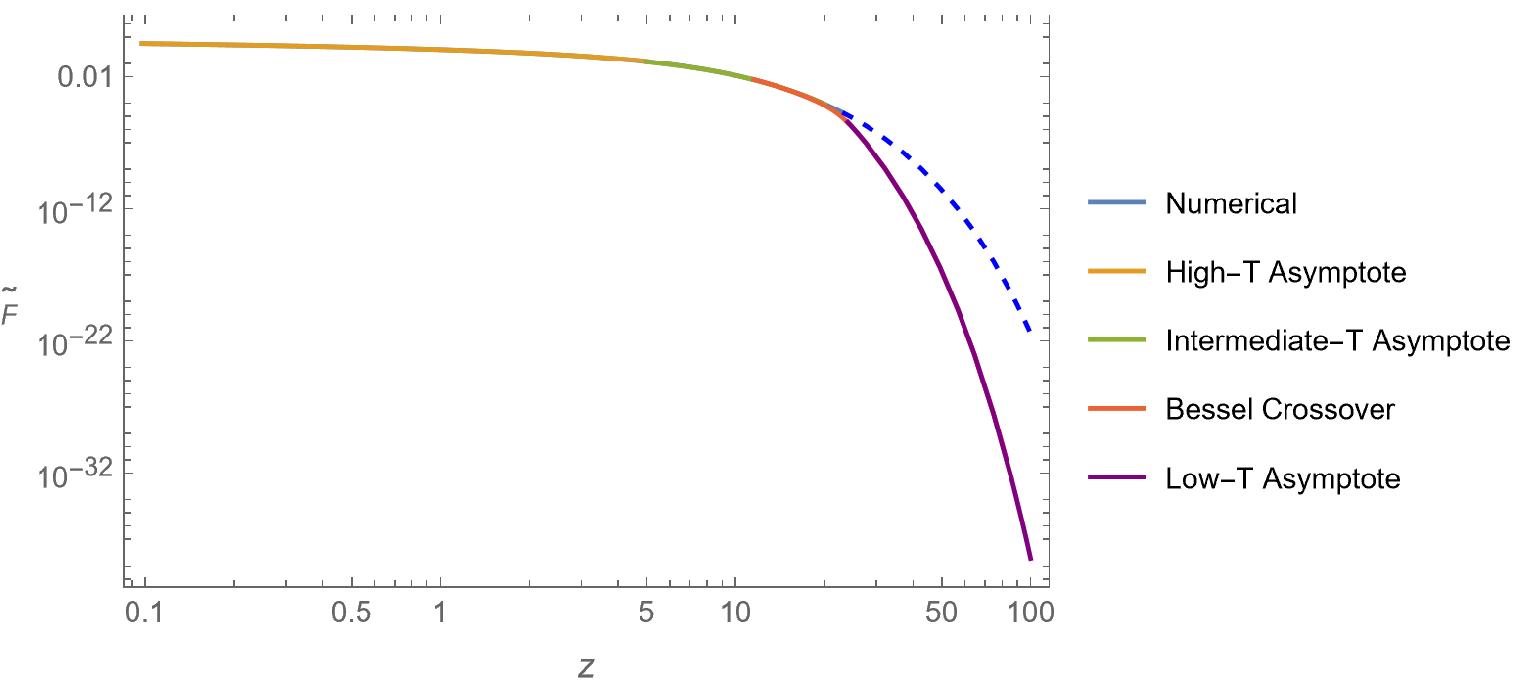}
  \caption{Reduced free energy $\widetilde{F}$ --
    exact (blue and red) and asymptotic expressions plotted against reciprocal temperature
    $z$ with $N=10^5$ (note, logarithmic scale on both axes).}
    \label{SummaryF}
\end{figure}

Equations (\ref{Fsmallz}) and (\ref{Flargez}) together with
eq.(\ref{FreeLow}) give the combined asymptotic formulae for the free energy
$F$ of the vortex gas in the \textit{High}, \textit{Intermediate}
and \textit{Low Temperature} regimes,
\begin{numcases}
{F=}
-NT\left[1-\log
  z+\frac{z}{6}-\frac{z^2}{360}+O(z^3)\right]\qquad &
\text{for $0<z<5$} \nonumber \\
-NT[2e^{-\frac{1}{2}z}+(1-z)e^{-z}+O(e^{-\frac{3}{2}z})] & \text{for
  $5<z<2\log N$} \nonumber \\
-NT\left[Ne^{-z}-\frac{1}{4}N^3e^{-2z}+O(N^5e^{-3z}) \right]
& \text{for $z>2\log N$} \,.
\label{Freeall}
\end{numcases}
$F$ decays exponentially when $z>5$, and decays even faster when
$z>2\log N$ and $F$ is no longer extensive. Recall also that $F$ can
be computed exactly. When $z<2\log N$, we solve (\ref{zx_0rel})
numerically and use formulae (\ref{Gdef}) and (\ref{Free}); when
$z>\log N$, we use formula (\ref{FreeBessel}). The expressions for the
reduced free energy $\widetilde{F}=F/(-NT)$ are all shown in Fig. \ref{SummaryF}.
\noindent
The exact expression for $\widetilde{F}$ almost completely overlaps
its asymptotes. The dashed blue curve is an extension of the
solid blue curve; for $N$ larger than $10^5$, the crossover occurs
further to the right. 

\begin{figure}[H]
    \centering
    \includegraphics[width=14cm]{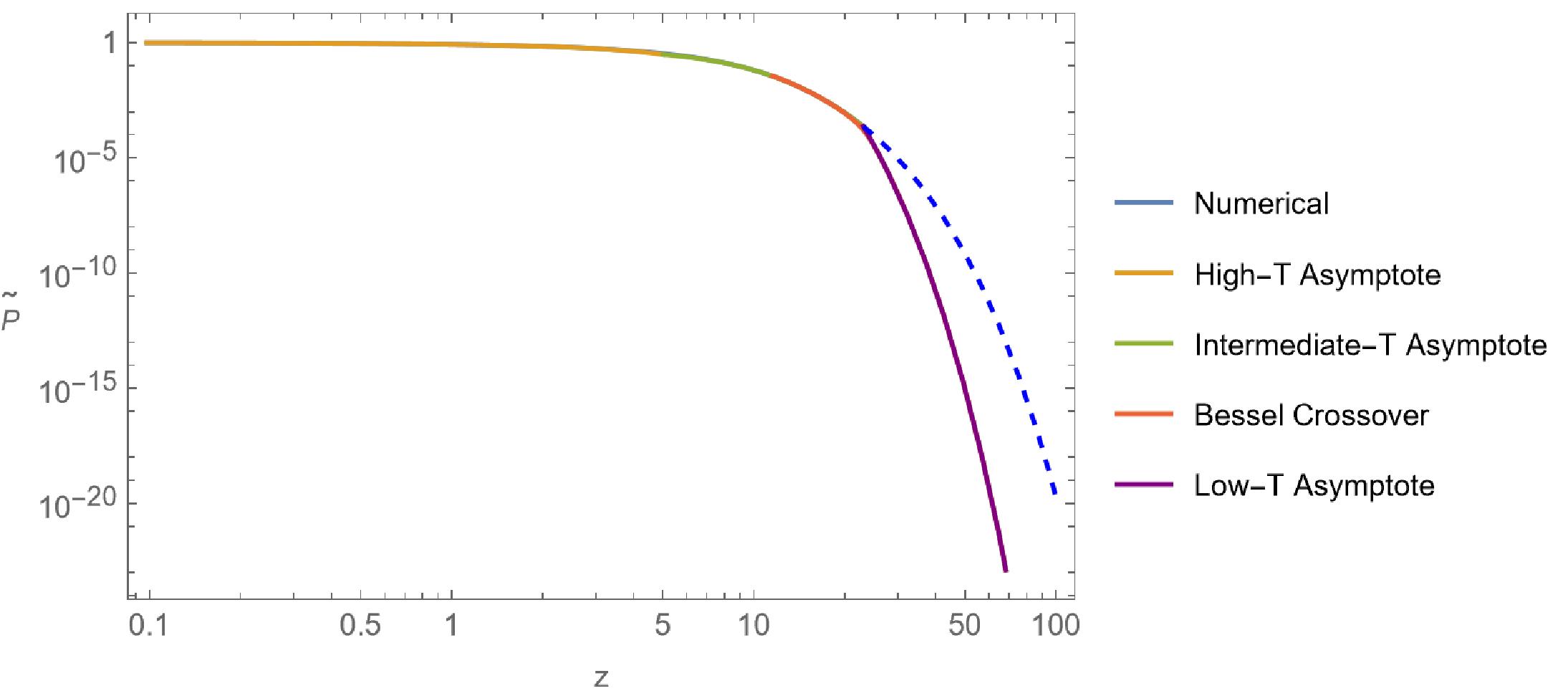}
    \caption{Reduced pressure $\widetilde{P}$ --
      exact (blue and red) and asymptotic results plotted against
      reciprocal temperature $z$, for $N=10^5$.}
    \label{SummaryP}
  \end{figure}

Equations (\ref{PHigh}), (\ref{PIntermediate}) and (\ref{PLow}) give
the combined asymptotic formulae for the pressure $P$ in the \textit{High},
\textit{Intermediate} and \textit{Low Temperature} regimes,
\begin{numcases}
{P=}
\frac{NT}{A-4\pi
  N}\left[1-\frac{z}{6}+\frac{z^2}{180}+O(z^3)\right]\qquad
& \text{for $0<z<5$} \nonumber \\
\frac{NT}{A-4\pi
  N}[ze^{-\frac{1}{2}z}+z(2-z)e^{-z}+O(e^{-\frac{3}{2}z}))] & \text{for
  $5<z<2\log N$} \nonumber \\
\frac{NT}{A-4\pi N}[Nze^{-z}-N^3ze^{-2z}+O(N^5ze^{-3z})] &
\text{for $z>2\log N$} \,.
\label{Pressureall}
\end{numcases}
The reduced pressure $\widetilde{P}=P\,/\,(\frac{NT}{A-4\pi N})$,
the quantity in square brackets, can also be computed exactly; when
$z<2\log N$, we solve (\ref{zx_0rel}) numerically and use (\ref{Px_0z}), and
when $z>\log N$, we use formula (\ref{PBessel}). These results are
shown in Fig. \ref{SummaryP} -- the numerically obtained solid blue curve is again
hardly visible. As before, the dashed blue curve extends the blue
curve, and the crossover occurs further to the right if $N$ is
larger than $10^5$.

\section{Conclusions}

We have considered a quantized gas of $N$ critically coupled Abelian
Higgs vortices on a surface diffeomorphic to $S^2$ having an area
close to the Bradlow limit, where the vortex density is just below
$\frac{1}{4\pi}$ and the surface is almost fully covered
by dissolving vortices. We have computed the partition function for
all (reciprocal) temperatures, and have derived the
free energy and pressure of the vortex gas. The calculations use the
known eigenvalues and degeneracies of the Beltrami Laplacian on
the $N$-vortex moduli space, which for dissolving vortices is
$\mathbb{CP}^N$ with a scaled Fubini--Study metric. Our results extend
the high temperature formulae for the free energy $F$ and pressure $P$
in~\cite{Man22} to all temperatures, but they apply to a more
limited range of vortex densities.

It is striking that $F\propto N^2$ at very low temperatures, below the
phase crossover from the \textit{Intermediate} to the
\textit{Low Temperature} regime, and it would be interesting to
better understand this non-extensiveness.

\begin{figure}[H]
  \centering
  \includegraphics[width=12cm]{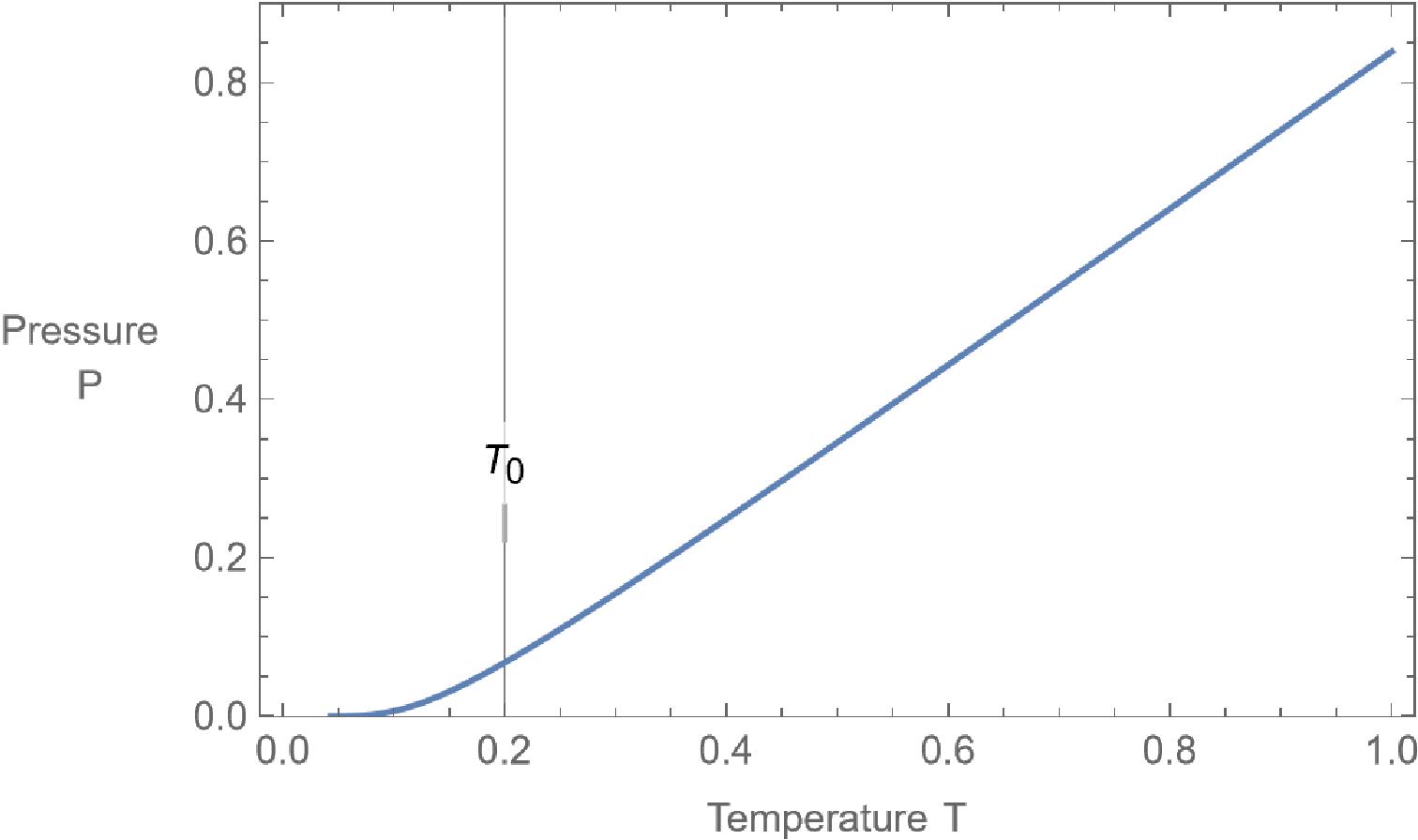}
  \caption{Scaled pressure $P$ against scaled temperature
    $T$; $T_0$ corresponds to reciprocal temperature $z=5$.}
  \label{ActualPT}
\end{figure}

In the \textit{Intermediate Temperature} regime, the pressure decays
exponentially with an exponent inversely proportional to temperature.
This is shown in the $T<T_0$ region of Fig. \ref{ActualPT}, where scaled pressure is plotted against
scaled temperature. One may compare this behaviour with an ideal
Bose gas, where the pressure has a rapid power law decrease as the
temperature approaches zero.

\section*{Acknowledgement}
S.W. is grateful for generous support from St John's College.


\begin{thebibliography}{99}

\bibitem{Bra} S.~B. Bradlow, 
Vortices in holomorphic line bundles over closed K\"ahler manifolds,
\textit{Commun. Math. Phys.} \textbf{135}, 1 (1990).

\bibitem{book} N. Manton and P. Sutcliffe,
\textit{Topological Solitons}, Cambridge Monographs on Mathematical Physics,
Cambridge University Press, 2004.

\bibitem{MN} N.~S. Manton and S.~M. Nasir,
Volume of vortex moduli spaces,
\textit{Commun. Math. Phys.} \textbf{199}, 591 (1999).
 
\bibitem{Bap} J.~M. Baptista,
On the $L^2$-metric of vortex moduli spaces,
\textit{Nucl. Phys.} \textbf{B844}, 308 (2011).

\bibitem{Man} N.~S. Manton,
Statistical mechanics of \small vortices, \textit{Nucl. Phys.} \textbf{B400} [FS], 624 (1993).
\normalsize

\bibitem{Man22} N.~S. Manton, 
Quantum statistical mechanics of vortices,
\textit{J. Phys. A: Math. Theor.} \textbf{55}, 325001 (2022).

\bibitem{BM} J.~M. Baptista and N.~S. Manton,
The dynamics of vortices on $S^2$ near the Bradlow limit,
\textit{J. Math. Phys.} \textbf{44}, 3495 (2003).

\bibitem{GS} R.~I. Garc\'ia Lara and J.~M. Speight,
The geometry of the space of vortices on a two-sphere in the Bradlow
limit, arXiv:2210.00966 (2022).

\bibitem{BGM} M. Berger, P. Gauduchon and E. Mazet,
\textit{Le Spectre d'une Vari\'et\'e Riemannienne}, Lecture Notes in
Math. \textbf{194}, Springer, Berlin, Heidelberg, 1971.

\end{thebibliography}
\end{document}